\documentclass[times,final]{elsarticle}
\usepackage{framed,multirow}

\usepackage{amssymb}
\usepackage{amsmath,mathtools,amsfonts,stmaryrd,bm}
\usepackage{latexsym}

\usepackage[left=2.5cm, right=2.5cm, top=3cm, bottom = 3cm]{geometry}

\usepackage{fancyhdr}

\usepackage{url}
\usepackage{xcolor}
\definecolor{newcolor}{rgb}{.8,.349,.1}

\definecolor{T}{rgb}{1,0,0}

\definecolor{O}{rgb}{0,0.7,0}

\newcommand{\overbar}[1]{\mkern 1.5mu\overline{\mkern-1.5mu#1\mkern-1.5mu}\mkern 1.5mu}
\newcommand{\Ub}{\ensuremath{\overbar{U}}}

\newcommand{\wb}{\ensuremath{\overbar{w}}}
\newcommand{\hb}{\ensuremath{\overbar{h}}}
\newcommand{\bb}{\ensuremath{\overbar{b}}}
\newcommand{\hub}{\ensuremath{\overbar{hu}}}

\newcommand{\Uh}{\ensuremath{\hat{U}}}
\newcommand{\wh}{\ensuremath{\hat{w}}}
\newcommand{\hh}{\ensuremath{\hat{h}}}
\newcommand{\bh}{\ensuremath{\hat{b}}}
\newcommand{\huh}{\ensuremath{\widehat{hu}}}

\newcommand{\realn}{\ensuremath{\mathbb{R}^n}}

\journal{arxiv}

\begin{document}

%\verso{T. Kent \textit{etal}}

\begin{frontmatter}

\title{Ensuring `well-balanced' shallow water flows via a discontinuous Galerkin finite element method: issues at lowest order}%
% \tnotetext[tnote1]{JCP short note: 4 pages or less (including figures, tables, and references but excluding title pages) and should not have an abstract.}

\author[1]{Thomas Kent\corref{cor1}}
\cortext[cor1]{Author correspondence:
t.kent@leeds.ac.uk; o.bokhove@leeds.ac.uk}
% \author[1]{Onno \snm{Bokhove}\fnref{fn1}}
% \fntext[fn1]{This is author footnote for second author.}  
\author[1]{Onno Bokhove}
%\author[1]{Steven \snm{Tobias}}

\address[1]{School of Mathematics and Leeds Institute of Fluid Dynamics, University of Leeds, Leeds LS2 9JT, UK.}

% \received{1 May 20XX}
% \finalform{10 May 20XX}
% \accepted{13 May 20XX}
% \availableonline{15 May 20XX}
% \communicated{S. Sarkar}

\begin{abstract}
% %%%
The discontinuous Galerkin finite element method (DGFEM) developed by \citet{rhe2008} offers a robust method for solving systems of nonconservative hyperbolic partial differential equations but, as we show here, does not satisfactorily deal with topography in shallow water flows at lowest order (so-called DG0, or equivalently finite volume). In particular, numerical solutions of the space-DG0 discretised one-dimensional shallow water equations over varying topography are not truly `well-balanced'.
%\footnote{This issue was reported first in \cite{ken2016}.}. 
A numerical scheme is well-balanced if trivial steady states are satisfied in the numerical solution; in the case of the shallow water equations, initialised rest flow should remain at rest for all times. Whilst the free-surface height and momentum remain constant and zero, respectively, suggesting that the scheme is indeed well-balanced, the fluid depth and topography evolve in time. This is both undesirable and unphysical, leading to incorrect numerical solutions for the fluid depth, and is thus a concern from a predictive modelling perspective.
We expose this unsatisfactory issue, both analytically and numerically, and indicate a solution that combines the DGFEM formulation for nonconservative products with a fast and stable well-balanced finite-volume method. This combined scheme bypasses the offending issue and successfully integrates nonconservative hyperbolic shallow water-type models with varying topography at lowest order. We briefly discuss implications for the definition of a well-balanced scheme, and highlight applications when higher-order schemes may not be desired, which give further value to our finding beyond its exposure alone.
% %%%%
\end{abstract}

% \begin{keyword}
% % Keywords
% \KWD discontinuous Galerkin finite element methods \sep finite volume method \sep nonconservative products \sep hyperbolic partial differential equations  \sep shallow water flows
% \end{keyword}

\end{frontmatter}


\noindent\hrulefill

%\linenumbers
%% main text
%%%%%%%%%%%%%%%%%%%%%%%%%%%%%%%%%%%%%%%%%%%%%%%%%%%%%%%%%%%%%%%%%%%%%%%%%%%%%%%%%%%%%%%%%%%%%%
\section{Introduction}
%%%%%%%%%%%%%%%%%%%%%%%%%%%%%%%%%%%%%%%%%%%%%%%%%%%%%%%%%%%%%%%%%%%%%%%%%%%%%%%%%%%%%%%%%%%%%%
Shallow water flows are ubiquitous in nature and engineering; their governing equations -- the shallow water equations (SWEs) -- form a hyperbolic system of partial differential equations (PDEs) and have a rich research history from both an analytical and numerical perspective (cf.~\citet{zei2018}). 
%The shallow water equations (SWEs) are considered a useful model in geophysical fluid dynamics for understanding the Earth’s atmosphere and oceans. 
There exists a powerful class of numerical methods for solving hyperbolic problems (e.g., ~\citet{lev2002}), often motivated by the need to capture shock formation which are a consequence of nonlinearities in the governing equations and manifest as discontinuities  in the solutions. One such numerical scheme that can be applied to hyperbolic problems is the discontinuous Galerkin finite element method (DGFEM); the main aims of this work are (i) to highlight an unsatisfactory issue of the DGFEM scheme developed by \citet{rhe2008} (hereon RBV2008), which concerns well-balancedness and arises when integrating the SWEs with varying topography at lowest order; and (ii) to give a comprehensive proof of the numerical artefact that causes it. Knowledge of this issue -- overlooked in RBV2008 and hitherto unreported in detail -- first arose in \citet{ken2017}, who commented on the problem but did not provide proof of the result. 
%In particular, numerical solutions of the space-DG0 discretised one-dimensional shallow water equations over varying topography are not truly `well-balanced'.
%\footnote{This issue was reported first in \cite{ken2016}.}. 
%A numerical scheme is well-balanced if trivial steady states are satisfied in the numerical solution; in the case of the SWEs, initialised rest flow should remain at rest for all times. Whilst the free-surface height and velocity remain constant and zero, respectively, suggesting that the scheme is indeed well-balanced, the fluid depth and topography evolve in time. We expose this unsatisfactory issue, overlooked in RBV2008, both analytically and numerically, and indicate a solution reported by \citet{ken2017} that combines the DGFEM methodology of RBV2008 with the scheme of \citet{aud2004} to integrate nonconservative hyperbolic shallow water-type models with varying topography at lowest order.
In order to elucidate the `well-balanced' issue in a consistent and concise manner, we outline the relevant background from RBV2008 in section 2 and then investigate in section 3 the relevant rest flow conditions at lowest order via analytical calculations and numerical simulations. We conclude in section 4 with a summary of the main result and a discussion of its implications.
%%%%%%%%%%%%%%%%%%%%%%%%%%%%%%%%%%%%%%%%%%%%%%%%%%%%%%%%%%%%%%%%%%%%%%%%%%%%%%%%%%%%%%%%%%%%%%
\section{1D DGFEM for nonconservative hyperbolic PDEs}\label{sec:dgfem2}
%%%%%%%%%%%%%%%%%%%%%%%%%%%%%%%%%%%%%%%%%%%%%%%%%%%%%%%%%%%%%%%%%%%%%%%%%%%%%%%%%%%%%%%%%%%%%%
In order to elucidate the `well-balanced' issue in a consistent and concise manner, we outline next the relevant background from RBV2008 and then investigate the rest flow conditions at lowest order. In particular, we recall briefly the DGFEM weak formulation for solving nonconservative hyperbolic systems of PDEs, i.e., systems of the form
\begin{equation}\label{eq:noncon3}
 \partial_{t} \mathbf{U} + \partial_{x} \pmb{F}(\pmb{U}) + \pmb{G}(\pmb{U}) \partial_x \pmb{U} = 0,
\end{equation}
where $\pmb{U} \in \realn$ are the model variables, $\pmb{F} \in \realn$ is a flux function and $\pmb{G} \in \realn \times \realn$ is the matrix of nonconservative products (NCPs). Partial derivatives with respect to time $t$ and space $x$ are denoted by $ \partial_{t}$ and $ \partial_{x}$ respectively. Since the system is hyperbolic, the Jacobian $\partial \pmb{F} / \partial \pmb{U} + \pmb{G} \in \realn \times \realn$ has real eigenvalues. It is non-conservative in the sense that $\pmb{G}(\pmb{U}) \partial_x \pmb{U}$ cannot be expressed in terms of a flux function $\partial_{x} \widetilde{\pmb{F}}(\pmb{U})$, i.e., there is no function $\widetilde{\pmb{F}}$ such that $\partial_{\pmb{U}} \widetilde{\pmb{F}} = \pmb{G}$. 
% Alternatively, the system (\ref{eq:noncon3}) can be written:
% \begin{equation}\label{eq:noncon2}
%  \partial_{t} \pmb{U} + \pmb{D}(\pmb{U}) \partial_x \pmb{U} = 0,
% \end{equation}
% where $\pmb{D} = \partial \pmb{F} / \partial \pmb{U} + \pmb{G} \in \realnn$.
Crucial to the weak formulation derived for equations of the form (\ref{eq:noncon3}) is DLM theory \citep{mas1995}, which regularizes the problem to overcome the absence of a weak solution (when the solution becomes discontinuous) due to the nonconservative products $\pmb{G}(\pmb{U}) \partial_x \pmb{U}$ \citep{rhe2008}.

%%%%%%%%%%%%%%%%%%%%%%%%%%%%%%%%%%%%%%%%%%%%%%%%%%%%%%%%%%%%%%%%%%%%%%%%%%%%%%%%%%%%%%%%%%%%%%
\subsection{Weak formulation and discretization}\label{sec:weakf}
%%%%%%%%%%%%%%%%%%%%%%%%%%%%%%%%%%%%%%%%%%%%%%%%%%%%%%%%%%%%%%%%%%%%%%%%%%%%%%%%%%%%%%%%%%%%%%
The one-dimensional domain $\Omega = {[0,L]}$ is divided into $N_{el}$ elements $K_k = (x_{k} , x_{k+1})$ for $k=1, 2, ..., N_{el}$ with $N_{el}+1$ nodes/edges $x_{1}, x_{2}, ..., x_{N_{el}}, x_{N_{el}+1}$. Element lengths $|K_k| = x_{k+1} - x_{k}$ may vary. Formally, after RBV2008, we define a tessellation $\mathcal{T}_h$ of the $N_{el}$ elements $K_k$:
\begin{equation}\label{eq:mesh}
\mathcal{T}_h = \left\{ K_k : \bigcup_{k=1}^{N_{el}} \bar{K}_k = \bar{\Omega}, K_k \cap K_{k'} = \emptyset \text { if }  k \ne k', 1 \le k,k' \le N_{el} \right\},
\end{equation}
where the overbar denotes closure $\bar{\Omega} = \Omega \cup \partial \Omega$, i.e., the elements $K_k$ cover the whole domain and do not overlap. Computational states are generally continuous on each element but discontinuous at the nodes.
The space DGFEM weak formulation for the system (\ref{eq:noncon3}) is given by equation (A 11) in RBV2008 and reproduced here in Eq.~(\ref{eq:wfd}). Repeated indices are used for the summation convention with $i,j = 1,...,n$ denoting components of vectors; $k$-subscript denotes values in element $K_k$; $L,R$-superscripts and $+,-$-superscripts denote limiting functional values and $x$ values, respectively, to the left/right of an element edge.
% (e.g., $U_k^R = U(x_{k}^{R})  = \lim_{x \downarrow x_{k}} U(x,t)$, and $U_k^L = U(x_{k}^{L})  = \lim_{x \uparrow x_{k}} U(x,t)$). 
In one space dimension and considering cell $K_k$ only, the weak form reads:
\begin{align}\label{eq:wfd}
0 = \int_{K_k} \left[ w \partial_{t} U_i -  F_i \partial_x w + w G_{ij} \partial_x U_j \right] \mathrm{d}x %\nonumber \\
% &\quad 
+ \left[ w (x_{k+1}^{-}) \mathcal{P}_i^p (x_{k+1}^{-}, x_{k+1}^{+})- w (x_{k}^{+}) \mathcal{P}_i^m (x_{k}^{-}, x_{k}^{+}) \right],
\end{align}
where $\mathcal{P}^p$ and $\mathcal{P}^m$ are given by:
% \begin{subequations}%\label{eq:pppm}
\begin{align}\label{eq:pppm}
\mathcal{P}_i^p = \hat{\mathcal{P}}_i^{NC} +\frac{1}{2} \int_0^1 G_{ij} (\phi) \frac{\partial \phi_j}{\partial \tau} \mathrm{d}\tau, \quad
\mathcal{P}_i^m = \hat{\mathcal{P}}_i^{NC} -\frac{1}{2} \int_0^1 G_{ij} (\phi) \frac{\partial \phi_j}{\partial \tau} \mathrm{d}\tau,
\end{align}
% \end{subequations}
and the NCP flux is:
\begin{align}\label{eq:NCP}
\hat{\mathcal{P}}_i^{NC}  (\pmb{U}^L, \pmb{U}^R) = 
\begin{cases}
F_i^L - \frac{1}{2} \int_0^1 G_{ij} (\phi) \frac{\partial \phi_j}{\partial \tau} \mathrm{d}\tau, &\quad \text{ if } S^L>0;\\
F_i^{HLL} - \frac{1}{2} \frac{S^L + S^R}{S^R-S^L} \int_0^1 G_{ij} (\phi) \frac{\partial \phi_j}{\partial \tau} \mathrm{d}\tau, &\quad \text{ if } S^L<0<S^R;\\
F_i^R + \frac{1}{2} \int_0^1 G_{ij} (\phi) \frac{\partial \phi_j}{\partial \tau} \mathrm{d}\tau, &\quad \text{ if } S^R<0.    
\end{cases}
\end{align}
% This is achieved by considering a single NCP $g(u) \partial_x u$, where $g$ is a smooth function but $u$ may admit discontinuities, and defining a smooth regularization $u^\epsilon$ of the discontinuous $u$:
% \begin{equation}\label{eq:reg}
%   g(u) \frac{\mathrm{d}u}{\mathrm{d}x} \equiv \lim_{\epsilon\rightarrow 0}  g( u^\epsilon ) \frac{\mathrm{d}u^\epsilon}{\mathrm{d}x}  = C \delta_{x_d}, \text{ with } C = \int_0^1 g(\phi(\tau)) \frac{\partial \phi}{\partial \tau} (\tau) \mathrm{d}\tau,
% \end{equation}
In the above integrals, $\phi:[0,1] \rightarrow \realn$ is a Lipschitz continuous path, satisfying $\phi(0) = \pmb{U}^L$ and $\phi(1) = \pmb{U}^R$, and connects the model states across the discontinuities arising naturally at the element boundaries in the DGFEM framework\footnote{This path is an artefact of the regularization of an NCP via DLM theory \citep{mas1995}, discussed briefly in \citet{ken2017} and in more detail in \citet{rhe2008}.}. Finally, $F_i^{HLL}$ is the standard HLL numerical flux \citep{har1983}
\begin{equation}\label{eq:hll}
 F_i^{HLL} = \frac{F_i^L S^R - F_i^R S^L + S^L S^R (U_i^R - U_i^L)}{S^R - S^L},
\end{equation}
$G_{ij}$ is the $ij$-th element of the matrix $\pmb{G}$, and $S^{L,R}$ are the fastest left- and right-moving signal velocities in the solution of the Riemann problem, determined by the eigenvalues of the Jacobian $\partial \pmb{F} / \partial \pmb{U} + \pmb{G}$ of the system. 
% %%%%%%%%%%%%%%%%%%%%%%%%%%%%%%%%%%%%%%%%%%%%%%%%%%%%%%%%%%%%%%%%%%%%%%%%%%%%%%%%%%%%%%%%%%%%%%
% \subsection{Space-DG0 discretization}
% %%%%%%%%%%%%%%%%%%%%%%%%%%%%%%%%%%%%%%%%%%%%%%%%%%%%%%%%%%%%%%%%%%%%%%%%%%%%%%%%%%%%%%%%%%%%%%
% \noindent Using piecewise constant basis functions, we take $U \approx U_h = \Ub_k (t)$ and, since the test function $w \approx w_h$ is arbitrary, $w_h = 1$ alternately in each element. The semi-discrete space-DGFEM scheme for element $K_k$ reads:
% \begin{equation}\label{eq:scheme0}
% 0 = |K_k| \frac{\mathrm{d} \Ub_k}{\mathrm{d}t} + \mathcal{P}^p (\Ub_{k+1}^{-}, \Ub_{k+1}^{+}) - \mathcal{P}^m (\Ub_{k}^{-}, \Ub_{k}^{+}),
% \end{equation}
% where $\Ub_{k+1}^{-} = \Ub_{k}$, $\Ub_{k+1}^{+} = \Ub_{k+1}$, $\Ub_{k}^{-} = \Ub_{k-1}$, $\Ub_{k}^{+} = \Ub_{k}$. This is analagous to a `Finite Volume' (FV) Godunov scheme in one dimension.

%%%%%%%%%%%%%%%%%%%%%%%%%%%%%%%%%%%%%%%%%%%%%%%%%%%%%%%%%%%%%%%%%%%%%%%%%%%%%%%%%%%%%%%%%%%%%%
\section{Does rest flow remain at rest?}\label{sec:balance}
%%%%%%%%%%%%%%%%%%%%%%%%%%%%%%%%%%%%%%%%%%%%%%%%%%%%%%%%%%%%%%%%%%%%%%%%%%%%%%%%%%%%%%%%%%%%%%
The topography $b$ in a shallow water model can be treated as a model variable ($b=b(x,t)$ with $\partial_t b = 0$) such that the nonconservative topographic term $-gh \partial_x b$ is then treated as an NCP. To highlight the issue of well-balanced flows, we consider the non-rotating shallow water equations with non-zero bottom topography:
\begin{subequations}\label{eq:SWEsNCP}
\begin{align}
&\partial_{t} h + \partial_{x} (hu) = 0, \label{eq:h}\\
&\partial_{t} (hu) + \partial_{x} \left( hu^2 + \frac{1}{2}gh^2 \right) = - gh \partial_{x}b, \label{eq:hu}\\
&\partial_{t} b = 0 \label{eq:b},
\end{align}
\end{subequations}
 which can be expressed in non-conservative form (\ref{eq:noncon3}) with:
\begin{equation}\label{eq:SWEs2}
  \pmb{U} =
  \begin{bmatrix}
  h\\hu\\b
  \end{bmatrix}, \quad
 \pmb{F}(\pmb{U}) =
  \begin{bmatrix}
  hu\\hu^2 + \frac{1}{2}gh^2\\0
  \end{bmatrix}, \quad
   \pmb{G}(\pmb{U}) =
  \begin{bmatrix}
  0 & 0 & 0 \\ 
  0 & 0 & gh\\ 
  0 & 0 & 0 
  \end{bmatrix}.
\end{equation}
The eigenvalues of the Jacobian $\partial \pmb{F} / \partial \pmb{U} + \pmb{G}$ are $\lambda_{\pm} = u \pm \sqrt{gh}$ and $\lambda_0 = 0$, which give the following numerical speeds:
\begin{subequations}
\begin{align}
S^L = \mathrm{min}\left( u^L - \sqrt{gh^L}, u^R - \sqrt{gh^R} \right)\quad\textrm{and}\quad
%,\\
S^R = \mathrm{max}\left( u^L + \sqrt{gh^L}, u^R + \sqrt{gh^R} \right).
\end{align}
\end{subequations}
For $i=1,3$, there are no NCPs in the equations so contributions to the integrals in (\ref{eq:pppm}) and (\ref{eq:NCP}) are zero. For $i=2$ and after employing a linear path ${\phi}(\tau;\pmb{U}^L,\pmb{U}^R) = \pmb{U}^L + \tau(\pmb{U}^R - \pmb{U}^L)$, one finds that:
\begin{align}\label{eq:vnc2}
\int_0^1 G_{2j} ({\phi}) \frac{\partial \phi_j}{\partial \tau} \mathrm{d}\tau &= \int_0^1g(h^L + \tau (h^R - h^L))(b^R - b^L) \mathrm{d}\tau = -g  \llbracket b \rrbracket \{\!\{ h \}\!\},
\end{align}
% \begin{align}\label{eq:vnc2}
% \int_0^1 G_{2j} (\pmb{\phi}) \frac{\partial \phi_j}{\partial \tau} \mathrm{d}\tau &= \int_0^1g(h^L + \tau (h^R - h^L))(b^R - b^L) \mathrm{d}\tau \nonumber\\
% &= g (b^R - b^L) \int_0^1(h^L + \tau (h^R - h^L)) \mathrm{d}\tau \nonumber\\
% &= g (b^R - b^L) \left[ h^L \tau + \frac{1}{2}\tau^2 (h^r - h^L) \right]_0^1  \nonumber\\
% &= g (b^R - b^L) \frac{1}{2} (h^L+ h^R)  \nonumber\\
% &= -g  \llbracket b \rrbracket \{\!\{ h \}\!\},
% \end{align}
where $\{\!\{ \cdot \}\!\} = \frac{1}{2}((\cdot)^L + (\cdot)^R)$ and $\llbracket \cdot \rrbracket = (\cdot)^L - (\cdot)^R$.
It is shown analytically here that when taking a linear path\footnote{Note that these calculations (Eqs.~\ref{eq:vnc2} -- \ref{eq:bevol}) hold when taking a $n$-polynomial path ${\phi}(\tau;\pmb{U}^L,\pmb{U}^R) = \pmb{U}^L + \tau^n(\pmb{U}^R - \pmb{U}^L)$. In fact, the main result (Eq.~\ref{eq:bevol}) is independent of ${\phi}$; see appendix B.} and lowest order (DG0, i.e., piecewise constant) approximation for the model states and test functions, the resulting scheme is not truly well-balanced.
% \textcolor{red}{[However, if you take another path; does this remain true; the unbalancedness? I.e. by taking another path can you rederive (21) and show you can or cannot keep db/dt=0?]}
Flow at rest requires that the free surface height remains constant $b^L + h^L = b^R + h^R$ with $u^L = u^R = 0$. Under these conditions, $S^L<0<S^R$ always and so the NCP flux (\ref{eq:NCP}) is:
\begin{equation}
\hat{\mathcal{P}}_i^{NC} = F_i^{HLL} - \frac{1}{2} \frac{S^L + S^R}{S^R-S^L} V_i^{NC},
\end{equation}
where $V_i^{NC} = \int_0^1 G_{ij} ({\phi}) \frac{\partial \phi_j}{\partial \tau} \mathrm{d}\tau$ is zero for $i=1,3$ and given by (\ref{eq:vnc2}) for $i = 2$. Since $F_1 = hu = 0$ for rest flow and $F_3 = 0$, the fluxes for the $h$- \eqref{eq:h} and $b$-equations \eqref{eq:b} are:
\begin{equation}
\hat{\mathcal{P}}_1^{NC} = \frac{S^L S^R (h^R - h^L)}{S^R - S^L}, \quad \hat{\mathcal{P}}_3^{NC} = \frac{S^L S^R (b^R - b^L)}{S^R - S^L}.
\end{equation}
%and by (\ref{eq:pppm}) we have that $\mathcal{P}_1^p = \mathcal{P}_1^m = \hat{\mathcal{P}}_1^{NC}$ and $\mathcal{P}_3^p = \mathcal{P}_3^m = \hat{\mathcal{P}}_3^{NC}$. 
For the $hu$-equation, we note that under rest flow conditions $U_2 = hu = 0$, $F_2 = \frac{1}{2}g h^2$, $V_2^{NC} = \frac{1}{2} g((h^L)^2 - (h^R)^2)$, and $\llbracket b \rrbracket = b^L - b^R = h^R - h^L$; thus, the second component of the NCP flux is $\hat{\mathcal{P}}_2^{NC} =  \frac{1}{4} g ((h^L)^2 + (h^R)^2)$.
%\begin{align}
%\hat{\mathcal{P}}_2^{NC} 
%%&=\frac{1}{S^R-S^L} \Big[ S^R F_2^L - S^L F_2^R  -\frac{1}{2} (S^L + S^R) V_2^{NC} \Big] \nonumber \\
%%&= \frac{1}{S^R-S^L} \Big[ \frac{1}{2} S^R g (h^L)^2 - \frac{1}{2} S^L g (h^R)^2  -\frac{1}{4} (S^L + S^R) g ((h^L)^2 - (h^R)^2)\Big] \nonumber \\
%% &=  \frac{1}{4} g \frac{1}{S^R-S^L} \Big[ S^R (h^L)^2 - S^L (h^R)^2 - S^L (h^L)^2 + S^R (h^R)^2)\Big] \nonumber \\
%% &=  \frac{1}{4} g \frac{1}{S^R-S^L} \Big[ (S^R - S^L) ((h^L)^2 + (h^R)^2)\Big] \nonumber \\
%&=  \frac{1}{4} g ((h^L)^2 + (h^R)^2). 
%%\\
%%\implies \mathcal{P}_2^{p} &= \hat{\mathcal{P}}_2^{NC}  + \frac{1}{2} V_2^{NC} = \frac{1}{4} g ((h^L)^2 + (h^R)^2) + \frac{1}{4} g((h^L)^2 - (h^R)^2) 
%%= \frac{1}{2} g (h^L)^2, \\
%%\implies \mathcal{P}_2^{m} &= \hat{\mathcal{P}}_2^{NC}  - \frac{1}{2} V_2^{NC} =  \frac{1}{4} g ((h^L)^2 + (h^R)^2) - \frac{1}{4} g((h^L)^2 - (h^R)^2) = \frac{1}{2} g (h^R)^2.
%\end{align}
The flux functions in Eq.~\ref{eq:pppm} are therefore:
\begin{equation}\label{eq:Pflux}
\mathcal{P}^{p} =
  \begin{bmatrix}
 \frac{S^L S^R (h^R - h^L)}{S^R - S^L} \\ \frac{1}{2} g (h^L)^2 \\ \frac{S^L S^R (b^R - b^L)}{S^R - S^L}
  \end{bmatrix}, \quad
\mathcal{P}^{m} =
  \begin{bmatrix}
 \frac{S^L S^R (h^R - h^L)}{S^R - S^L} \\ \frac{1}{2} g (h^R)^2 \\ \frac{S^L S^R (b^R - b^L)}{S^R - S^L}
  \end{bmatrix}.
\end{equation}
Following RBV2008, but using piecewise constant basis functions $w \approx w_h = 1$ alternately in each element and $U \approx U_h = \Ub_k (t)$, the space-DG0 (finite volume) scheme for element $K_k$ reads:
\begin{equation}\label{eq:scheme}
0 = |K_k| \frac{\mathrm{d} \Ub_k}{\mathrm{d}t} + \mathcal{P}^p (\Ub_{k+1}^{-}, \Ub_{k+1}^{+}) - \mathcal{P}^m (\Ub_{k}^{-}, \Ub_{k}^{+}),
\end{equation}
where left- and right-states $\Ub_{k+1}^{-} = \Ub_{k}$, $\Ub_{k+1}^{+} = \Ub_{k+1}$, $\Ub_{k}^{-} = \Ub_{k-1}$, $\Ub_{k}^{+} = \Ub_{k}$ yield numerical fluxes: 
\begin{equation}
\mathcal{P}^{p} =
  \begin{bmatrix}
 \frac{S_{k+1}^L S_{k+1}^R (\hb_{k+1} - \hb_{k})}{S_{k+1}^R - S_{k+1}^L} \\ \frac{1}{2} g \hb_k^2 \\ \frac{S_{k+1}^L S_{k+1}^R (\bb_{k+1} - \bb_{k})}{S_{k+1}^R - S_{k+1}^L}
  \end{bmatrix}, \quad
\mathcal{P}^{m} =
  \begin{bmatrix}
 \frac{S_k^L S_k^R (\hb_k - \hb_{k-1})}{S_k^R - S_k^L} \\ \frac{1}{2} g \hb_k^2 \\ \frac{S_k^L S_k^R (\bb_k - \bb_{k-1})}{S_k^R - S_k^L}
  \end{bmatrix}.
\end{equation}
Conditions for rest flow are assessed by considering the evolution of momentum $hu$ and free surface height $h+b$, as determined by the DG0 discretization (\ref{eq:scheme}): 
%\textcolor{red}{[align below better; on less lines?]}:
%\begin{align}
%hu &: \quad  0 = |K_k| \frac{\mathrm{d}}{\mathrm{d}t} (\hub_k) +  \frac{1}{2} g \hb_k^2 -  \frac{1}{2} g \hb_k^2 \implies \frac{\mathrm{d}}{\mathrm{d}t} (\hub_k)  = 0, \\
%h+b &: \quad  0 = |K_k| \frac{\mathrm{d}}{\mathrm{d}t} (\hb_k + \bb_k) + \frac{S_{k+1}^L S_{k+1}^R (\hb_{k+1} - \hb_{k} + \bb_{k+1} - \bb_{k})}{S_{k+1}^R - S_{k+1}^L}
%%\nonumber \\ &\qquad \qquad 
% -  \frac{S_k^L S_k^R (\hb_k - \hb_{k-1} + \bb_k - \bb_{k-1})}{S_k^R - S_k^L}\nonumber\\ &\qquad \implies \frac{\mathrm{d}}{\mathrm{d}t}  (\hb_k + \bb_k)   = 0,
%\end{align}
\begin{subequations}\label{eq:restflow}
\begin{align}
&0 = |K_k| \frac{\mathrm{d}}{\mathrm{d}t} (\hub_k) +  \frac{1}{2} g \hb_k^2 -  \frac{1}{2} g \hb_k^2 \implies \frac{\mathrm{d}}{\mathrm{d}t} (\hub_k)  = 0, \\
&0 = |K_k| \frac{\mathrm{d}}{\mathrm{d}t} (\hb_k + \bb_k) + \frac{S_{k+1}^L S_{k+1}^R (\hb_{k+1} - \hb_{k} + \bb_{k+1} - \bb_{k})}{S_{k+1}^R - S_{k+1}^L}
 -  \frac{S_k^L S_k^R (\hb_k - \hb_{k-1} + \bb_k - \bb_{k-1})}{S_k^R - S_k^L} \implies \frac{\mathrm{d}}{\mathrm{d}t}  (\hb_k + \bb_k)   = 0,
\end{align}
\end{subequations}
since $h^L + b^L = h^R + b^R$. Thus, both $h+b$ and $hu$ remain constant when initialised with rest flow, and the scheme appears to be well-balanced.
%%%%%%%%%%%%%%%%%%%%%%%%%%%%%%%%%%%%%%%%%%%%%%%%%%%%%%%%%%%%%%%%%%%%%%%%%%%%%%%%%%%%%%%%%%%%%%
% FIGURE
%%%%%%%%%%%%%%%%%%%%%%%%%%%%%%%%%%%%%%%%%%%%%%%%%%%%%%%%%%%%%%%%%%%%%%%%%%%%%%%%%%%%%%%%%%%%%%
\begin{figure}[t]
\centering
\includegraphics[width = \textwidth]{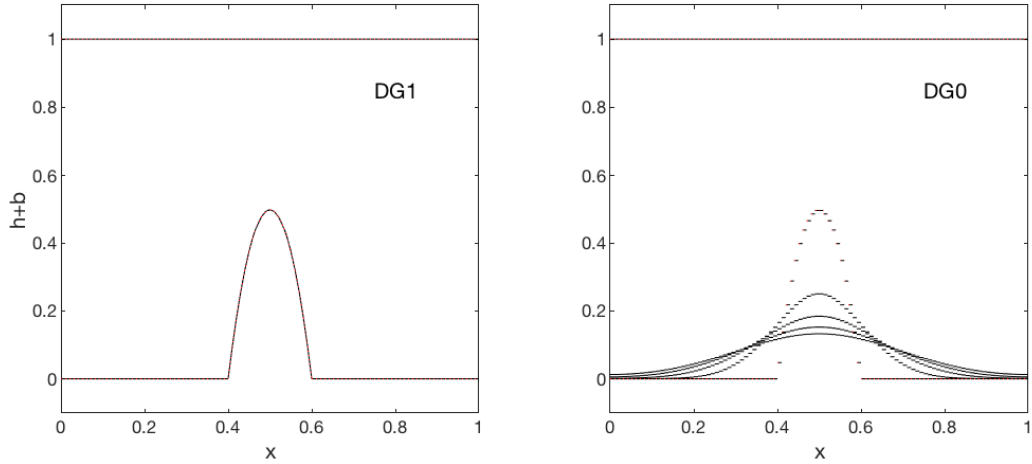}
\caption{Snapshots of the free-surface height $h+b$ and topography $b$ at times at $t=0, 2.5,5,7.5, 10$ in DG1 (piecewise linear; left) and DG0 (piecewise constant; right) simulations initialised with rest flow conditions ($h+b=1$ and $hu=0$). DG1 simulations, with piecewise linear topography continuous across elements, maintain flow at rest for all $t>0$ and are therefore considered well-balanced. On the other hand, evolving topography (and therefore fluid depth $h$) emerges as $t>0$ in DG0 simulations. Despite $h+b=1$ for all $t$ at DG0, the evolving $b$ and $h$ means that the scheme should not be considered truly well-balanced. Other simulation details: $N_{el} = 100$, $\mathrm{Fr} = 1.9$ (supercritical), topography given in Eq.\ref{eq:bic}, $t = [0,10]$.}
\label{fig:fig1}
\end{figure}
%%%%%%%%%%%%%%%%%%%%%%%%%%%%%%%%%%%%%%%%%%%%%%%%%%%%%%%%%%%%%%%%%%%%%%%%%%%%%%%%%%%%%%%%%%%%%%
However, consider the evolution of $b$ only:
\begin{align}\label{eq:bevol}
0 = |K_k| \frac{\mathrm{d}}{\mathrm{d}t} (\bb_k) + \frac{S_{k+1}^L S_{k+1}^R (\bb_{k+1} - \bb_{k})}{S_{k+1}^R - S_{k+1}^L} - \frac{S_{k}^L S_{k}^R (\bb_{k} - \bb_{k-1})}{S_{k}^R - S_{k}^L},
\end{align}
and note that the evolution equation for $h$ is the same as Eq.~\eqref{eq:bevol} after replacing $b$ with $h$ everywhere. These equations for the numerical integration of $b$ and $h$ are the crux of this article and it is here that our analysis goes further than RBV2008 to expose the following issue. Since $b \approx b_h$ is discontinuous at the nodes for non-constant $b$, the sum of the flux terms is non-zero (for $S_{k}^{L,R} \ne 0$), leading to non-steady topography; the same is true for $h$. Thus, although flow remains at rest in the sense that $h+b=const.$ and $hu=0$, the DG0 scheme is not truly well-balanced in the sense that $\mathrm{d}\bb_k / \mathrm{d}t  \ne 0$ and $\mathrm{d} \hb_k/ \mathrm{d}t \ne 0$. That is, the fluid depth $h$ and topography $b$ evolve in time. This peculiar artefact is demonstrated numerically in Fig.~1, which shows both DG0 and DG1 simulations initialised with rest flow conditions and integrated with a standard third-order Runge-Kutta time-step routine. We solve the non-dimensionalized equations, effectively setting $g = \mathrm{Fr}^{-2}$ in Eq.\ref{eq:SWEsNCP} where $\mathrm{Fr}$ is the Froude number, on a domain of length $L=1$. For topography, the classical profile of an isolated parabolic ridge is implemented (as in RBV2008):
\begin{align}\label{eq:bic}
b(x,t) = 
\begin{cases}
b_c \left( 1 - \left( \frac{x - x_p}{a}\right)^2 \right), &\quad \text{ for } |x-x_p| \le a; \\
0, &\quad \text{ otherwise,}
\end{cases}
\end{align}
where $b_c$ is the height of the hill crest, $a$ is the hill width parameter, and $x_p$ its location in the domain. For DG0 (right panel of Fig.~1) simulations, the analytical evolution of $b$ in Eq.~\eqref{eq:bevol} is exemplified for this set-up. Despite the free-surface height $h+b$ remaining constant, the topography $b$ `diffuses' as $t>0$ so that the fluid depth $h$ is also non-steady. 
For DG1 expansions (and higher-order), we can project the DG expansion coefficients of $b$ such that $b_h$ remains continuous across elements, then $b^R = b^L$ and $\mathrm{d}\bb_k / \mathrm{d}t = 0$. Then all aspects of rest flow are satisfied numerically and the scheme can be considered truly well-balanced (see left panel of Fig.~1). We note also that, for DG1 and above, if $b_h$ is initially discontinuous across elements (cf.~\cite{rhe2008}) then it evolves to a nearby continuous solution, at which point numerical solutions for $b$ and $h$ remain steady.
For completeness, a proof that the DG1 discretization satisfies all aspects of well-balanced flow, first published in RBV2008, is reproduced here using our notation in appendix A. 
%Further comments on the effect of the regularization path are given in appendix B.

\section{Conclusion}
This short article has shown that modelling topography as a time-independent variable in a shallow water system (Eq.~\ref{eq:SWEsNCP}) and solving via the DGFEM scheme of RBV2008 is not suitable at lowest order (DG0, or finite volume). Whilst the conditions for rest flow are apparently satisfied (Eq.~\ref{eq:restflow}), an unsatisfactory artefact of the DG0 discretization, namely unsteady water depth and topography, is exposed analytically in Eq.~(\ref{eq:bevol}) and numerically in Fig.~\ref{fig:fig1}. This is both undesirable and unphysical, leading to incorrect solutions for $h$ and $b$, and is thus a concern from a predictive modelling perspective. This result also highlights that a well-balanced scheme (in the sense of satisfying $h+b=const.$ and $hu=0$ only) can lead to unsteady fluid depth $h$, which is wholly inadequate; perhaps to be considered truly well-balanced, a scheme must satisfy further conditions that $h$ and $b$ are steady in separation.

We recognise that there are other higher-order DG schemes reported in the literature that do not suffer from this issue, including RBV2008, but note that such schemes may not always be desirable when computational cost is a major consideration (cf.~\citet{ken2017}). In weather forecasting, for example, higher order accuracy may need to be sacrificed for gains in computational efficiency, especially when real-time forecasting is combined with real-time data assimilation.
The modified shallow water model of \citet{ken2017} includes not only topography but also other nonconservative terms relating to idealized atmospheric convection. Motivated by the need to provide a computationally inexpensive solver for data assimilation research, a low-order (i.e., DG0) discretization in \citet{ken2017} was of greater importance than improved accuracy of higher order discretizations. 
The hitherto unforeseen issue detailed here, and first encountered in  \citet{ken2017}, has been bypassed by combining the theory of RBV2008 for dealing with the NCPs and the method of \citet{aud2004} for dealing with the topography. The resulting scheme successfully integrates nonconservative hyperbolic shallow water-type models with varying topography at lowest order.

\section*{Acknowledgments}
TK acknowledges support from the Engineering and Physical Sciences Research Council and the Met Office [grant number 1305398]. We thank Prof.~Steven Tobias and Dr.~Gordon Inverarity for helpful discussions.

% \section*{References}

% Please ensure that every reference cited in the text is also present in
% the reference list (and vice versa).

% \section*{\itshape Reference style}

% Text: All citations in the text should refer to:
% \begin{enumerate}
% \item Single author: the author's name (without initials, unless there
% is ambiguity) and the year of publication;
% \item Two authors: both authors' names and the year of publication;
% \item Three or more authors: first author's name followed by `et al.'
% and the year of publication.
% \end{enumerate}
% Citations may be made directly (or parenthetically). Groups of
% references should be listed first alphabetically, then chronologically.

%%Vancouver style references.
\bibliographystyle{model1-num-names}
\bibliography{refs}

\appendix
\section{DG1 discretization}

The DG1 discretization uses piecewise linear basis functions (i.e., first-order polynomials) to approximate the trial function $U$ and test function $w$ and thereby discretise the weak formulation (\ref{eq:wfd}) in space. The DG1 expansions are
\begin{equation}
U \approx U_h = \Ub + \xi \Uh; \quad w \approx w_h = \wb +\xi \wh.
\end{equation}
with mean and slope coefficients $\Ub = \Ub_k (t)$ and $\Uh = \Uh_k (t)$, where $\xi \in (-1,1)$ is a local coordinate in the reference element $\hat{K}_k$ such that:
\begin{equation}
x = x(\xi) = \frac{1}{2} \big( x_k + x_{k+1} + |\hat{K}_k| \xi \big).
\end{equation}
Thus, when $\xi = -1$, $x = x_k$ and $\xi = 1$, $x = x_{k+1}$. Also note that $\mathrm{d}x  = \frac{1}{2} |\hat{K}_k|\mathrm{d}\xi$. We evaluate the integrals in (\ref{eq:wfd}) with $w_i = w_i |_{K_k}$ and $U_i = U_i |_{K_k}$ as follows:
%\textcolor{red}{[Align below better --less lines?]}:
\begin{align}\label{eq:int1}
\int_{K_k} w_i \partial_{t} U_i \mathrm{d}x  &= \int_{K_k} (\wb_i +\xi \wh_i) \partial_{t} (\Ub_i + \xi \Uh_i) \mathrm{d}x \nonumber \\
&= \frac{1}{2} |K_k| \int_{-1}^1 \wb_i \partial_{t} \Ub_i +(\wh_i \partial_{t} \Ub_i + \wb_i \partial_t \Uh_i)\xi + (\wh_i \partial_t \Uh_i ) \xi^2 \mathrm{d}\xi \nonumber \\
&= \frac{1}{2} |K_k| \left[ 2\wb_i \partial_{t} \Ub_i + \frac{2}{3}\wh_i \partial_t \Uh_i \right]
%\nonumber \\&
= |K_k| \wb_i \partial_{t} \Ub_i + \frac{1}{3} |K_k| \wh_i \partial_t \Uh_i,
\end{align}
%%%%%%%%%%%%%%%%%%%%%%%%%%%%%%%%%%%%%%%%%%%%%%%%%%%%%%%%%%%%%%%%%%%%%%%%%%%%%%%%%%%%%%%%%%%%%%
\begin{align} \label{eq:F_INT}
\int_{K_k} -F_i \partial_x w_i \mathrm{d}x  &= -\int_{K_k} F_i (\Ub + \xi \Uh) \partial_x (\wb_i +\xi \wh_i) \mathrm{d}x  \nonumber \\
&=  -\int_{-1}^1 F_i (\Ub + \xi \Uh) \frac{2}{|K_k|} \partial_{\xi} (\wb_i +\xi \wh_i)\frac{1}{2} |K_k| \mathrm{d}\xi 
%\nonumber \\&
=  - \wh_i \int_{-1}^1 F_i (\Ub + \xi \Uh)\mathrm{d}\xi, 
\end{align}
%%%%%%%%%%%%%%%%%%%%%%%%%%%%%%%%%%%%%%%%%%%%%%%%%%%%%%%%%%%%%%%%%%%%%%%%%%%%%%%%%%%%%%%%%%%%%%
\begin{align}\label{eq:G_INT}
\int_{K_k} w_i G_{ij} \partial_x U_j \mathrm{d}x  &= \int_{K_k} (\wb_i +\xi \wh_i) G_{ij} (\Ub + \xi \Uh) \partial_x (\Ub_j + \xi \Uh_j) \mathrm{d}x  \nonumber \\
&=   \int_{-1}^1 (\wb_i +\xi \wh_i) G_{ij} (\Ub + \xi \Uh) \frac{2}{|K_k|} \partial_\xi (\Ub_j + \xi \Uh_j) \frac{1}{2}|K_k|\mathrm{d}\xi  \nonumber \\
&=  \int_{-1}^1 (\wb_i +\xi \wh_i) G_{ij} (\Ub + \xi \Uh) \Uh_j \mathrm{d}\xi 
%\nonumber \\&
= \wb_i  \int_{-1}^1 G_{ij} (\Ub + \xi \Uh) \Uh_j \mathrm{d}\xi +\wh_i \int_{-1}^1 \xi G_{ij} (\Ub + \xi \Uh) \Uh_j \mathrm{d}\xi.  
\end{align}
%In general, for numerical implementation, the integrals (\ref{eq:F_INT}, \ref{eq:G_INT}) are approximated with a Gauss quadrature rule, e.g.,  third-order two-point:
%\begin{equation}
%\int_{-1}^1 \psi(\xi) \mathrm{d}\xi \approx \psi(-c_m) + \psi(c_m)
%\end{equation}
%where $c_m = 1/\sqrt{3}$ for some function $\psi(\xi)$. 
The flux terms in (\ref{eq:wfd}) are:
\begin{equation}
w_i (x_{k+1}^{-}) \mathcal{P}_i^p (x_{k+1}^{-}, x_{k+1}^{+}) = (\wb_i + \wh_i) |_{K_k} \mathcal{P}_i^p \big((\Ub_i  + \Uh_i )|_{K_k}, (\Ub_i- \Uh_i )|_{K_{k+1}}\big),
\end{equation}
\begin{equation}
w_i (x_{k}^{+}) \mathcal{P}_i^m (x_{k}^{-}, x_{k}^{+}) = (\wb_i - \wh_i) |_{K_k} \mathcal{P}_i^m \big((\Ub_i  + \Uh_i )|_{K_{k-1}}, (\Ub_i - \Uh_i )|_{K_{k}}\big).
\end{equation}
The space-discretised scheme for means $\Ub_i$ and slopes $\Uh_i$ is obtained by considering coefficients of the test function means $\wb_i$ and slopes $\wh_i$ and taking $\wb_i = \wh_i = 1$ alternately for each element (again due to arbitrariness of $w_h$):
\begin{subequations}\label{eq:dwf_final}
\begin{align}
0 &= |K_k| \partial_{t} \Ub_i + \mathcal{P}_i^p \big(U^L|_{K_k}, U^R|_{K_{k+1}}\big) - \mathcal{P}_i^m \big(U^L|_{K_{k-1}}, U^R|_{K_{k}}\big) 
%\nonumber \\&\qquad 
+ \int_{-1}^1 G_{ij} (\Ub + \xi \Uh) \Uh_j \mathrm{d}\xi  \\
0 &= \frac{1}{3} |K_k| \partial_t \Uh_i + \mathcal{P}_i^p \big(U^L|_{K_k}, U^R|_{K_{k+1}}\big) + \mathcal{P}_i^m \big(U^L|_{K_{k-1}}, U^R|_{K_{k}}\big) 
%\nonumber \\&\qquad
- \int_{-1}^1 F_i (\Ub + \xi \Uh)\mathrm{d}\xi + \int_{-1}^1 \xi G_{ij} (\Ub + \xi \Uh) \Uh_j \mathrm{d}\xi,
\end{align}
\end{subequations}
where $U^L = \Ub + \Uh$ and $U^R = \Ub  - \Uh$ are the trace values to the left and right of a element edge.

 Here it is shown analytically that when taking a linear path and using first-order expansion for the model states and test functions, rest flow in the shallow water system (\ref{eq:SWEsNCP}) remains at rest and the non-constant topography $b$ does not evolve as long as $b_h$ remains continuous across elements.
The semi-discrete scheme is given by (\ref{eq:dwf_final}) and we evaluate the integrals therein for rest flow, and check the following:
\begin{equation}\label{eq:restcond}
 \frac{\mathrm{d}}{\mathrm{d}t}  (\hb_k + \bb_k)  = 0, \quad  \frac{\mathrm{d}}{\mathrm{d}t}  (\hh_k + \bh_k)  = 0, \quad  \frac{\mathrm{d}}{\mathrm{d}t}  (\hub_k)  = 0, \quad  \frac{\mathrm{d}}{\mathrm{d}t}  (\huh_k)  = 0.
\end{equation}
For $i=1,3$, integrals involving $G$ are zero. For $i=2$:
\begin{align}\label{eq:restG}
 &\int_{-1}^1 G_{2j} (\Ub + \xi \Uh) \Uh_j \mathrm{d}\xi  = g \int_{-1}^1 (\hb + \xi \hh) \bh \mathrm{d}\xi = g \int_{-1}^1 (\hb \bh +  \hh \bh \xi)\mathrm{d}\xi = 2 g \hb \bh,  \\
&\int_{-1}^1 \xi G_{2j} (\Ub + \xi \Uh) \Uh_j \mathrm{d}\xi = g \int_{-1}^1 \xi (\hb + \xi \hh) \bh \mathrm{d}\xi = g \int_{-1}^1  (\hb \bh \xi + \hh \bh \xi^2) \mathrm{d}\xi = \frac{2}{3} g \hh \bh
\end{align}
with the first integral featuring in the equation for means $\Ub_i$ and the second in the equation for slopes $\Uh$. For the integral involving the flux $F$:
\begin{subequations}\label{eq:restF}
\begin{align}
\int_{-1}^1 F_1 (\Ub + \xi \Uh)\mathrm{d}\xi &= \int_{-1}^1 (\hub + \xi \huh)\mathrm{d}\xi = 0, \text{ since flow is at rest;} \\
\int_{-1}^1 F_2 (\Ub + \xi \Uh)\mathrm{d}\xi &= \int_{-1}^1 \frac{1}{2}g (\hb +\xi \hh)^2 \mathrm{d}\xi = \frac{1}{2}g \int_{-1}^1  (\hb^2 +2\xi \hb\hh +\xi^2 \hh^2) \mathrm{d}\xi \nonumber \\
&= \frac{1}{2}g \left[ 2 \hb^2 +  \frac{2}{3}\hh^2 \right] =  g \hb^2 + \frac{1}{3} g \hh^2;  \\
\int_{-1}^1 F_3 (\Ub + \xi \Uh)\mathrm{d}\xi  &= 0. 
\end{align}
\end{subequations}
Using (\ref{eq:Pflux}), (\ref{eq:restG}), and (\ref{eq:restF}) in (\ref{eq:dwf_final}), we check the conditions (\ref{eq:restcond}) for rest flow to be satisfied numerically:
\begin{align}
\hb+\bb&: \quad 0 = |K_k| \frac{\mathrm{d}}{\mathrm{d}t} (\hb_k + \bb_k) + \frac{S_{k+1}^L S_{k+1}^R (\underline{\underline{h^R_{k+1} - h^L_{k+1} + b^R_{k+1} - b^L_{k+1}}})}{S_{k+1}^R - S_{k+1}^L} 
%\nonumber \\ &\qquad  \qquad 
 -  \frac{S_k^L S_k^R (\underline{\underline{h^R_k - h^L_{k} + b^R_k - b^L_{k}}})}{S_k^R - S_k^L} \nonumber \\
&\qquad\qquad\implies \frac{\mathrm{d}}{\mathrm{d}t}  (\hb_k + \bb_k)   = 0;
\end{align}
%%%%%%%%%%%%%%%%%%%%%%%%%%%%%%%%%%%%%%%%%%%%%%%%%%%%%%%%%%%%%%%%%%%%%%%%%%%%%%%%%%%%%%%%%%%%%%
\begin{align}
\hh+\bh&: \quad 0 = \frac{1}{3} |K_k| \frac{\mathrm{d}}{\mathrm{d}t} (\hh_k + \bh_k) + \frac{S_{k+1}^L S_{k+1}^R (\underline{\underline{h^R_{k+1} - h^L_{k+1} + b^R_{k+1} - b^L_{k+1}}})}{S_{k+1}^R - S_{k+1}^L} %\nonumber \\ &\qquad  \qquad 
+  \frac{S_k^L S_k^R (\underline{\underline{h^R_k - h^L_{k} + b^R_k - b^L_{k}}})}{S_k^R - S_k^L} \nonumber \\
&\qquad\qquad \implies \frac{\mathrm{d}}{\mathrm{d}t}  (\hh_k + \bh_k)   = 0; 
\end{align}
%%%%%%%%%%%%%%%%%%%%%%%%%%%%%%%%%%%%%%%%%%%%%%%%%%%%%%%%%%%%%%%%%%%%%%%%%%%%%%%%%%%%%%%%%%%%%%
\begin{align}
\hub&: \quad 0 = |K_k| \frac{\mathrm{d}}{\mathrm{d}t} (\hub_k) +  \frac{1}{2} g (\hb_k + \hh_k)^2 -  \frac{1}{2} g (\hb_k - \hh_k)^2 + 2 g \hb_k \bh_k 
%\nonumber \\&\qquad \quad 
= |K_k| \frac{\mathrm{d}}{\mathrm{d}t} (\hub_k) + 2 g \hb_k (\underline{\underline{\hh_k + \bh_k}}) \nonumber \\
&\qquad\qquad \implies \frac{\mathrm{d}}{\mathrm{d}t} (\hub_k)  = 0; 
\end{align}
%%%%%%%%%%%%%%%%%%%%%%%%%%%%%%%%%%%%%%%%%%%%%%%%%%%%%%%%%%%%%%%%%%%%%%%%%%%%%%%%%%%%%%%%%%%%%%
\begin{align}
\huh&: \quad 0 = \frac{1}{3}|K_k| \frac{\mathrm{d}}{\mathrm{d}t} (\huh_k) +  \frac{1}{2} g (\hb_k + \hh_k)^2 +  \frac{1}{2} g (\hb_k - \hh_k)^2 - g\hb_k^2 - \frac{1}{3} g \hh_k^2 + \frac{2}{3} g \hh_k \bh_k \nonumber \\
&\qquad \quad = \frac{1}{3} |K_k| \frac{\mathrm{d}}{\mathrm{d}t} (\huh_k) + g \hb_k^2 + g \hh_k^2 - g\hb_k^2 - \frac{1}{3} g \hh_k^2 + \frac{2}{3} g \hh_k \bh_k 
%\nonumber \\ &\qquad \quad
= \frac{1}{3} |K_k| \frac{\mathrm{d}}{\mathrm{d}t} (\huh_k) + \frac{2}{3} g \hh_k (\underline{\underline{\hh_k + \bh_k}}) \nonumber \\
&\qquad\qquad \implies \frac{\mathrm{d}}{\mathrm{d}t} (\huh_k)  = 0.
\end{align}
Twice-underlined terms in the above evaluations are zero after noting that, for flow at rest, $h^L + b^L = h^R + b^R$ and the slope of $h+b$ is zero. Thus, it has been proven that rest flow remains at rest for the DG1 space discretization when using a linear path. Moreover, if we consider the evolution of $b$ only:
\begin{subequations}
\begin{align}
0 &= |K_k| \frac{\mathrm{d}}{\mathrm{d}t} (\bb_k) + \frac{S_{k+1}^L S_{k+1}^R (b^R_{k+1} - b^L_{k+1})}{S_{k+1}^R - S_{k+1}^L} - \frac{S_{k}^L S_{k}^R (b^R_{k} - b^L_{k})}{S_{k}^R - S_{k}^L}  \\
0 &= \frac{1}{3} |K_k| \frac{\mathrm{d}}{\mathrm{d}t} (\bh_k) + \frac{S_{k+1}^L S_{k+1}^R (b^R_{k+1} - b^L_{k+1})}{S_{k+1}^R - S_{k+1}^L} + \frac{S_{k}^L S_{k}^R (b^R_{k} - b^L_{k})}{S_{k}^R - S_{k}^L}  
\end{align}
\end{subequations}
and project the topography $b$ such that $b_h$ remains continuous across elements (i.e., $b^R = b^L$), then $\mathrm{d}\bb_k/ \mathrm{d}t = \mathrm{d}\bh_k / \mathrm{d}t = 0$. Hence, all aspects of rest flow are satisfied numerically and the scheme is truly well-balanced. Note that, for DG1 and higher-order, if $b_h$ is initially discontinuous across elements then it evolves to a nearby continuous solution, at which point numerical solutions for $b$ and $h$ remain steady.

\section{Alternative path}

We employ a linear path ${\phi}$ to deal with the non-conservative products in the integral (\ref{eq:vnc2}). Consider instead an $n$-degree polynomial path:
\begin{align}
{\phi} ( \tau; \pmb{U}^L, \pmb{U}^R ) &= \pmb{U}^L + \tau^n ( \pmb{U}^R - \pmb{U}^L ), \quad \tau \in [0,1], 
%\\
\qquad \implies \frac{\partial {\phi}}{\partial \tau} = n \tau^{n-1} ( \pmb{U}^R - \pmb{U}^L ) = -n \tau^{n-1} \llbracket \pmb{U} \rrbracket .
\end{align}
Then the integral (\ref{eq:vnc2}) becomes:
\begin{align}\label{eq:vnc2n}
\int_0^1 G_{2j} ({\phi}) \frac{\partial \phi_j}{\partial \tau} \mathrm{d}\tau &= \int_0^1 g(h^L + \tau^n (h^R - h^L))n \tau^{n-1}(b^R - b^L) \mathrm{d}\tau 
%\nonumber\\&
= n g (b^R - b^L) \int_0^1 \tau^{n-1}(h^L + \tau^n (h^R - h^L)) \mathrm{d}\tau \nonumber\\
&= n g (b^R - b^L) \left[\frac{1}{n} h^L \tau^n + \frac{1}{2n}\tau^{2n} (h^R - h^L) \right]_0^1  %\nonumber\\&
= g (b^R - b^L) \frac{1}{2} (h^L+ h^R)  
%\nonumber\\&
= -g  \llbracket b \rrbracket \{\!\{ h \}\!\}.
\end{align}
Thus, taking a linear path does not affect the result. In fact, since there are no NCPs in the $h$- and $b$-evolution equations, the choice of path has no impact on the critical result (Eq.~\eqref{eq:bevol}).

\end{document}